\documentclass[aps,prl,twocolumn,superscriptaddress]{revtex4-1}
\usepackage{amsmath,amssymb,amsfonts,upgreek}
\usepackage{color}
\usepackage{graphicx}
\usepackage{setspace}
\usepackage{bm}
\usepackage[export]{adjustbox}
\usepackage{xr-hyper}
\usepackage{hyperref}
\usepackage{dsfont}
\usepackage{braket}
\usepackage{siunitx}
\usepackage{soul}

\usepackage{changes}

\newcommand{\hn}{\hat{n}}
\newcommand{\largeL}{$L\rightarrow\infty\,$}

\begin{document}

\title{Mott Transition and Volume Law Entanglement with Neural Quantum States}

\author{Chlo\'e Gauvin-Ndiaye}
\affiliation{D\'epartement de physique, Regroupement qu\'eb\'ecois sur les matériaux de pointe \& Institut quantique\\
Universit\'e de Sherbrooke, 2500 Boul. Universit\'e, Sherbrooke, Qu\'ebec J1K2R1, Canada}
\affiliation{Center for Computational Quantum Physics, Flatiron Institute, New York, New York 10010, USA}
\author{Joseph Tindall}
\affiliation{Center for Computational Quantum Physics, Flatiron Institute, New York, New York 10010, USA}
\author{Javier Robledo Moreno}
\affiliation{Center for Computational Quantum Physics, Flatiron Institute, New York, New York 10010, USA}
\affiliation{Center for Quantum Phenomena, Department of Physics, New York University, 726 Broadway, New York, New York 10003, USA}
\affiliation{IBM Quantum, IBM Research - 1101 Kitchawan Rd, Yorktown Heights, New York 10598, USA}
\author{Antoine Georges}
\affiliation{Coll\`{e}ge de France, PSL University, 11 place Marcelin Berthelot, 75005 Paris, France}
\affiliation{Center for Computational Quantum Physics, Flatiron Institute, New York, New York 10010, USA}
\affiliation{Department of Quantum Matter Physics, University of Geneva, 24 quai Ernest-Ansermet, 1211 Geneva, Switzerland}
\affiliation{CPHT, CNRS, \'{E}cole Polytechnique, IP Paris, F-91128 Palaiseau, France}

\date{\today}

\begin{abstract}

The interplay between delocalisation and repulsive interactions can cause electronic systems to undergo a Mott transition between a metal and an insulator. Here we use neural network hidden fermion determinantal states (HFDS) to uncover this transition in the disordered, fully-connected Hubbard model. Whilst dynamical mean-field theory (DMFT) provides exact solutions to physical observables of the model in the thermodynamic limit, our method allows us to directly access the wave function for finite system sizes well beyond the reach of exact diagonalisation. We demonstrate how HFDS are able to obtain more accurate results in the metallic regime and in the vicinity of the transition than calculations based on a Matrix Product State (MPS) ansatz, for which the volume law of entanglement exhibited by the system is prohibitive. 
We use the HFDS method to calculate the energy and double occupancy, the quasi-particle weight and the energy gap 
and, importantly, the amplitudes of the wave function which provide a novel insight into this model.
Our work paves the way for the study of strongly correlated electron systems with neural quantum states.

\end{abstract}

\maketitle

Neural quantum states (NQS) are transforming the field of variational wave functions for the 
quantum many-body problem. Originally introduced for 
models of interacting quantum spins \cite{CarleoNQS2017}, 
these methods have been generalized to interacting fermions 
for lattice models~\cite{inui_determinant-free_2021, stokes_phases_2020, nomura_restricted_2017, Nys2022variational, Luo2019backflow, Robledo_2022Hidden}, 
small molecular systems~\cite{xia2018_OG-NQS-QChem, choo_fermionic_2020, yang2020_NQSQChem, sureshbabu2021_NQSQC, Yoshioka2021solids, rath_framework_2023, barrett_autoregressive_2022, zhao_2022autoregressiveNQS} 
and systems in continuous space~\cite{ruggeri_nonlinear_2018, pfau_ab_2020, spencer_better_2020, Hermann2020Paulinet, Entwistle2023, Glehn2022Psiformer, Casella_2D_electron_gas_2023, pescia2023electronGas, kim2023fermiGas, lou2023FermiGas, Lovato_Hidden_nucleons_2022, Fore_Hidden_nucleons_2022}. 
%
Electron interactions can turn a metal into an insulator: this is the Mott phenomenon, which plays a 
central role in the physics of materials with strong electronic correlations. 
It is therefore relevant to ask whether NQS can describe both a correlated metallic phase and 
a Mott insulating phase, as well as the transition between them. 
The question is especially pertinent when the Mott insulating phase does not have long-range magnetic 
order in the ground-state, a notorious challenge 
to the expressivity of variational wave functions~\cite{Sorella2017QMCbible}.

Here, we consider one of the simplest models in which such a transition occurs: the fully-connected Hubbard model with random hopping amplitudes. This model is challenging for variational methods since, as shown below, it displays an entanglement entropy 
which scales with the number of sites. 
While successful in low-dimensional and tree-like systems \cite{Orus2019, tindall2020, tindall2023}, we find here that representing the wave function as a matrix product state (MPS) 
and optimizing it with the density-matrix renormalization group (DMRG) algorithm~\cite{White1992, White1993, White2005, Schollwoeck2011, Rommer1997, Jeckelmann2002} requires prohibitively large bond dimensions. 
Consequently, convergence can be achieved only up to moderate system sizes.

On the other hand, dynamical mean-field theory (DMFT)~\cite{Georges_1996}, 
which exploits locality rather than low entanglement, provides an exact framework in which the properties of this model in the limit 
of an infinite number of sites can be computed
with modern algorithms. 
This provides a controlled benchmark to compare variational methods to. 
Moreover, DMFT calculations for models such as the one studied 
here have played a fundamental part in our understanding
of the properties of strongly correlated materials and of the Mott transition\cite{Georges_1996,Kotliar_2006}. 
Being a Green's function based method, however, DMFT does not provide access to the many-body wave function 
of the system, which is hence still unknown
~\footnote{Note that while the Gutzwiller wave function is not the exact ground-state in the limit of large 
dimensions, recent extensions of the Gutzwiller {\it ansatz} provide improved variational wave functions in this limit 
~\cite{Lanata2023SlaveBoson, Lanata2017GhostGutzwiller}.}.

In this article, 
we study this model with the recently introduced neural network 
hidden fermion determinantal states (HFDS) \cite{Robledo_2022Hidden}. 
We show that, despite recent claims that NQS cannot represent physically relevant states that exhibit volume-law entanglement~\cite{Passetti_2023}, this method performs successfully both in the metallic and Mott insulating phases. 
It 
also clearly reveals the Mott transition, although accuracy is challenging in this regime. 
We show that while large-scale DMRG calculations working in the position basis also reveal the transition, for small interaction strengths and near the transition they incur a higher computational cost.
We also demonstrate that the performance of DMRG can be significantly improved by optimizing the basis of 
single-particle states.
Our work provides a new wave-function based perspective on this model, and 
paves the way to the study of physical phenomena associated with strong fermionic 
correlations with NQS.

\begin{figure}
    \centering
    \includegraphics[width=\columnwidth]{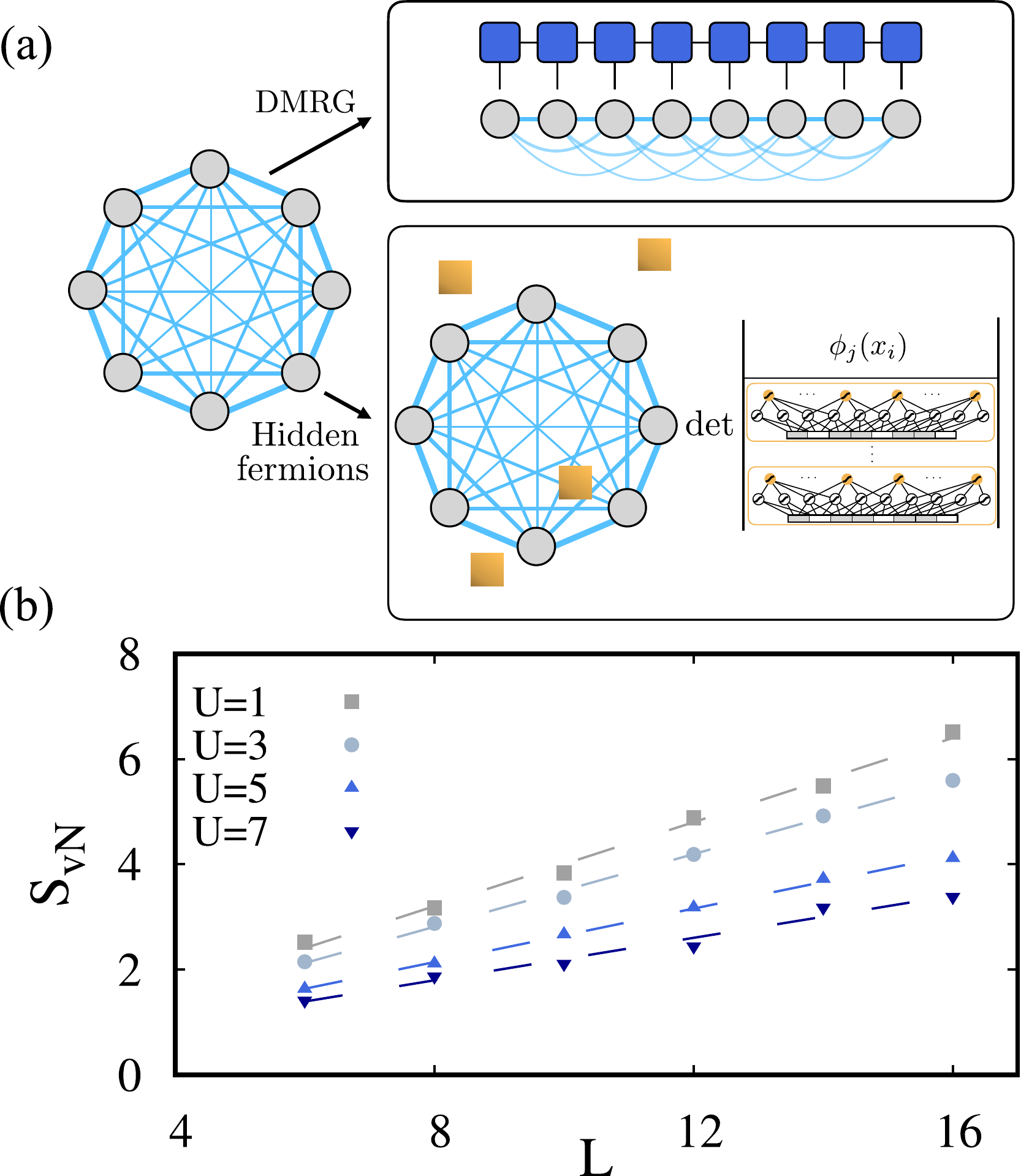}
   \caption{
     Panel (a): Schematic of the fully connected random hopping Hubbard model and the DMRG and HFDS methods. 
    Panel (b): Von-Neumann entanglement entropy of a random bipartition for the ground state of this model for various 
$U$ as a function of system size $L$, computed using DMRG with a maximum bond dimension of $32,000$. Dashed lines correspond to a linear fitting of the data. 
}
    \label{fig:1_Model_EE}
\end{figure}
\paragraph{Model.} - We consider the fully-connected fermionic Hubbard model of $L$ sites with random hopping amplitudes (Fig.~\ref{fig:1_Model_EE}) defined by the Hamiltonian: 
\begin{equation}
    H = -\frac{1}{\sqrt{L}}\sum_{\substack{i,j = 1 \\ i \neq j \\ }}^{L}
    \sum_{\sigma \in \{\uparrow, \downarrow \}}\,
    t_{ij}\,c^{\dagger}_{i, \sigma}c_{j, \sigma} + U \sum_{i=1}^{L} \hn_{i, \uparrow}\hn_{i, \downarrow},
    \label{Eq:Hamiltonian}
\end{equation}
where $c^{(\dagger)}_{i, \sigma}$ is the annihilation (creation) operator for a fermion of spin $\sigma$ on site $i$, and $\hn_{i, \sigma} = c^{\dagger}_{i, \sigma}c_{i, \sigma}$ is the number operator. 
The $L(L-1)/2$ hopping amplitudes $t_{ij}$ are random variables drawn independently from the Gaussian normal distribution 
with mean $0$ and standard deviation $1$, i.e. 
$\overline{t_{ij}}=0$ and $\overline{t^2_{ij}}=t^2=1$. We fix $t=1$ as our unit of energy. 
%
Throughout this work, we consider the half-filled case with, on average, one electron per site ($N=L$). 

In the infinite-volume limit, this model is known to possess {\it self-averaging} properties~\cite{SYK_review}. 
Local observables such as the double occupancy $d_i=\langle \hn_{i\uparrow}\hn_{i\downarrow}\rangle$ 
have sample-to-sample fluctuations which vanish as $L\rightarrow\infty$, converging with probability one to their sample-averaged values denoted by $\overline{\cdots}$ and which are site-independent: 
$d_i\rightarrow  d=\frac{1}{L}\sum_{i=1}^{L}\overline{\langle \hn_{i\uparrow}\hn_{i\downarrow}\rangle}$. For the wave function methods at finite $L$, we perform sample averages over $20$ realizations of the disorder for each value of $U$ and each system size. Unless stated otherwise, physical quantities shown in the plots are averaged over these realizations and the standard deviation is used as an estimation of the error bars. We use identical realizations for both DMRG and HFDS.

We now summarize the zero-temperature properties of the model known from DMFT 
for $L\rightarrow\infty$ (see the Supplemental Material (SM) \cite{supplemental}
 for an extended discussion). 
At finite $U$, the ground-state is a Fermi liquid up to a critical value $U=U_{c2}$. 
The most accurate solutions of the DMFT equations using the numerical renormalization group (NRG) impurity solver report $U_{c2}=5.82...$~\cite{Lee_2017}.  
For $U>U_{c2}$, the ground-state is a Mott insulator. For any finite $U$, however, charge fluctuations survive 
and the double occupancy remains nonzero. 

This model thus
provides one of the simplest examples of a system having a $T=0$ phase transition between 
a metallic Fermi liquid and a Mott insulator with no magnetic long-range order. 
The phase transition occurs at $U_{c2}$ and 
is second-order~\cite{Rozenberg_1994,Georges_1996}, 
the energy and double occupancy being continuous through the transition. 
A subtle point is that the insulating solution exists as a metastable, higher energy, solution in a window 
of coupling $U\in [U_{c1},U_{c2}]$, with $U_{c1}=4.74\cdots$~\cite{Lee_2017}.
This holds for the infinite-volume limit, but the consequences for finite-$L$ are not currently known. 

From a wave function point of view, this model is challenging since it has volume-law entanglement. 
In Fig.~\ref{fig:1_Model_EE}b, we evidence this: computing the half-system bi-partite von-Neumann 
entanglement entropy $S$ of the ground state 
as a function of $L$, up to $L = 16$, for a range of $U$. 
For all values of $U$ a clear linear dependence of $S$ with system size is observed -- suggesting that for any finite $U$ 
in the position basis the entanglement entropy follows a volume law. 
Recently, it has been argued that, despite being able to represent
artificially constructed entangled states, 
in practice deep and shallow feed-forward NQS are not capable of efficiently representing 
the ground states of physical models with volume-law entanglement~\cite{Passetti_2023}. 
Reference ~\cite{Passetti_2023}, however, does not consider functional parametrizations containing determinants, like Slater-Jastrow~\cite{nomura_restricted_2017, stokes_phases_2020}, backflow~\cite{Luo2019backflow} or HFDS~\cite{Robledo_2022Hidden} parametrizations. Determinant-based functional parametrizations play a fundamental role in the representability of the ground state of fermionic systems, regardless of the amount of entanglement. For example, the entanglement of the ground state of the Hamiltonian Eq.(\ref{Eq:Hamiltonian}) at $U=0$ follows a volume law in the localized basis. Despite this, it can be exactly represented by a Slater determinant. 

\paragraph{Methods.}
The HFDS formalism is based on an augmentation of the physical Fock space (spanned by $M = 2L$ `visible' fermionic degrees of freedom) by $\tilde{M}$ additional `hidden' modes, resulting in $M_\textrm{tot} = M + \tilde{M}$ degrees of freedom. 
The physical Fock space is recovered by a projection procedure, which assigns a unique state $\tilde{n}=F(n)$ of the hidden fermions 
to each Fock state $n=\{n_{i\sigma}\}$ ($n_{i\sigma}=0,1$) of the physical/visible fermions. 
The wave-function amplitude corresponding to the occupancy $n$ in the physical subspace is given by:
\begin{equation}\label{eq:HFS amplitudes}
    \psi(n) = \langle n, F(n) |\Psi \rangle,
\end{equation} where $|\Psi\rangle$ is a many-body state in the augmented space. 
A powerful variational state is obtained when both $|\Psi\rangle$ and $F(n)$ belong to a parametric family. 
In this work, following the construction in Ref.~\cite{Robledo_2022Hidden}, $|\Psi\rangle$ is taken to be a Slater determinant 
with $N$ visible and $\tilde{N}$ hidden fermions, and $F(n)$ is an internal state in a neural network. 
When projected onto the physical space, this yields a correlated wave function.
For lattice models, it can be shown that this ansatz is universal and is explicitly connected to configuration-interaction states~\cite{Robledo_2022Hidden, RobledoMorenoThesis2023}.
The amplitudes $\psi(n)$ are optimized using Variational Monte Carlo (VMC). 

The second method we use is DMRG~\cite{White1992,Schollwoeck2011}. 
We initialise our approximation for the ground state wave function as a product-state MPS
and repeatedly perform DMRG sweeps, enforcing a bond dimension $\chi \leq \chi_{\rm max}$. 
We utilize both $U(1)$ symmetries of the model -- restricting ourselves to half-filling and $ \langle S^{z} \rangle =0$. 
For system sizes $L \leq 16$ we use a sufficient bond-dimension ($\chi_{\rm max} = 32,000$) 
to observe convergence of the energy. For $U = 0$ and $L=16$, this means we reach an energy within $0.001 \%$ of the true ground-state energy. This is an extremely large bond dimension given the size of this system and is indicative of the difficulty of the problem at hand for the DMRG algorithm. 
For $L > 16$ and any $U$ we are unable to achieve convergence in energy and use the largest bond-dimension 
($\chi_{\rm max} \sim 10^{4}$) that we can afford with our available computational resources. 
Further details are provided in the SM ~\cite{supplemental}. 
There, we also show how DMRG calculations can be made more efficient by optimizing the 
basis of single-particle states, following the technique described in Ref.~\cite{Robledo2023BasisRotation}.

\paragraph{Results.}

\begin{figure}[t!]
    \centering
    \includegraphics[width=\columnwidth]{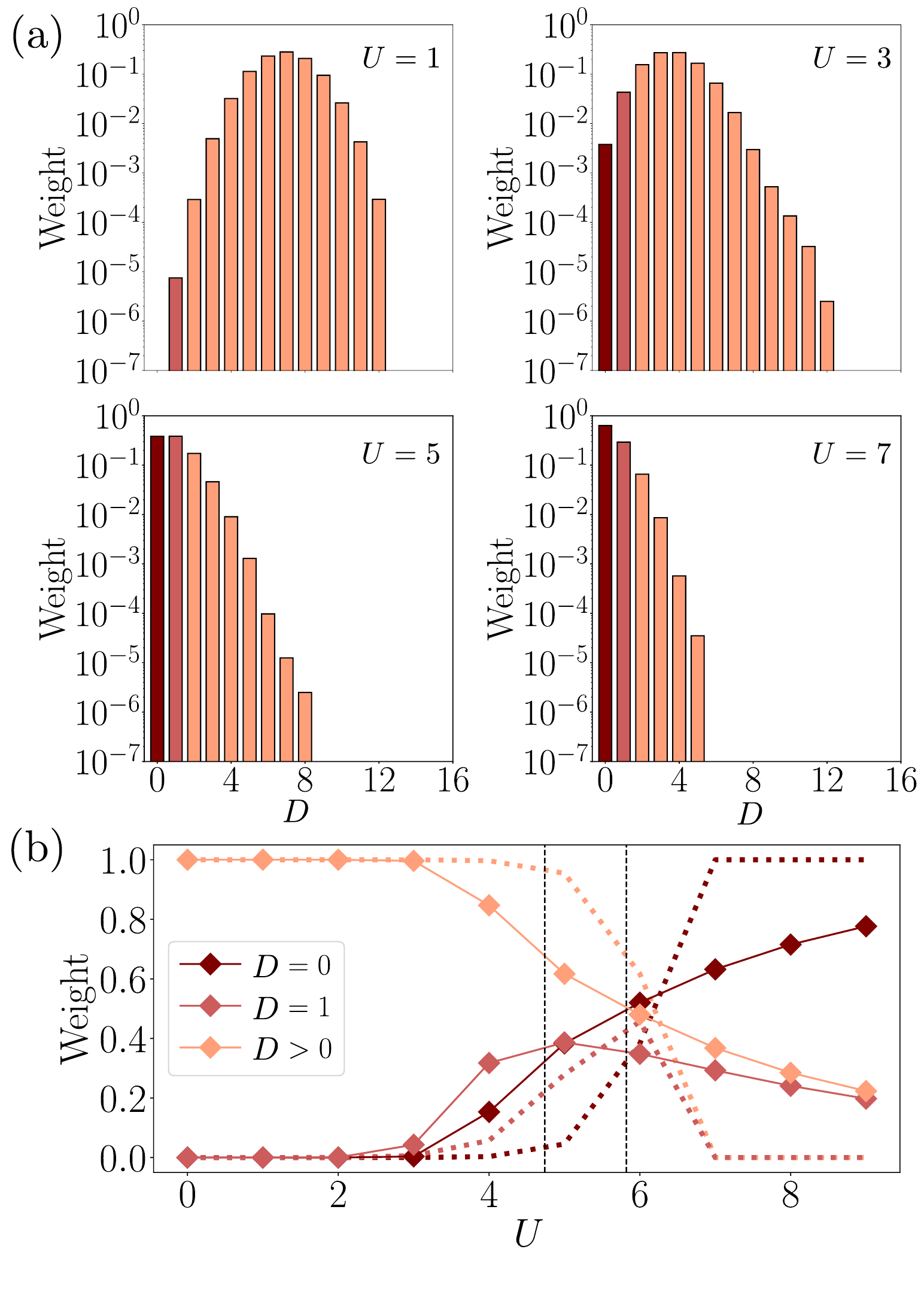}
    \caption{
    (a) Weight of the HFDS wave function for electronic configurations with double occupancy $D$ 
    (as defined in the main text), for $L=32$ and different values of $U$. Darker to lighter colors indicate $D = 0$, $D = 1$ and $D>0$ respectively, as related to panel (b). (b) Wave function weight versus $U$ for different values of the double occupancy as indicated. The dotted curves show the predictions from the Gutzwiller approach for reference (see ~\cite{supplemental}). 
    The dashed vertical lines indicate the values of $U_{c1}$ and $U_{c2}$ for $L=\infty$ as obtained 
    from DMFT. 
    }
    \label{fig:2_wave_function}
\end{figure}

To understand the structure of the wave function in the space of electronic configurations, we first study the weight of the wave-function amplitudes $\mathcal{W}(D) = \sum_{n_D} |\psi(n_D)|^2$ for electronic configurations $n_D$ 
which have a given number $D$ of doubly occupied sites. This quantity can be estimated as the ratio of sampled configurations with double occupancy $D$ when sampling Fock states $n_D$ according to the probability distribution $P(n_{d}) \propto |\psi(n_{D})|^2$. We provide a complete overview of this analysis in the End Matter.
Panel (a) in Fig.~\ref{fig:2_wave_function} shows a bar plot of these weights, 
obtained with the HFDS ansatz for a system with $L = 32$, which is beyond the reach of exact diagonalization. 
For small values of $U$ in the metallic phase, the wave function spreads across electronic configurations with a wide range of double occupancy. Close to the transition point, the weight shifts towards configurations with $D = 0$ and $D = 1$. For large values of $U$ in the Mott insulating phase, the weight of the wave function in each sub-sector of double occupancy decays exponentially with $D$ and peaks around $D = 0$. Panel (b) in Fig.~\ref{fig:2_wave_function} shows the weight of the wave function for $D = 0$, $D = 1$ and $D >0$ as a function of $U$. The point where the weight of the wave function becomes dominated by electronic configurations with $D = 0$ coincides with the metal-insulator transition, thus providing a signature of the phase transition. 
We also include, in panel (b) of  Fig.~\ref{fig:2_wave_function}, results from the Gutzwiller ansatz (GA)\cite{Vollhardt1984} 
(see ~\cite{supplemental} for calculation details and more data). These agree with our HFDS results deep in the metallic regime but the GA overestimates the critical point $U_{c}$ and, in its vicinity, predicts a larger weight for the $D= 0$ configurations than HFDS. This is a result of the GA falsely predicting that the charge fluctuations entirely 
vanish in the insulating phase. This observation is relevant to experiments: in the context of ultra cold atoms in optical lattices, small charge fluctuations in the Mott state have been observed in the form of residual interference \cite{Bloch2005}.

\begin{figure}[t!]
    \centering
    \includegraphics[width=\columnwidth]{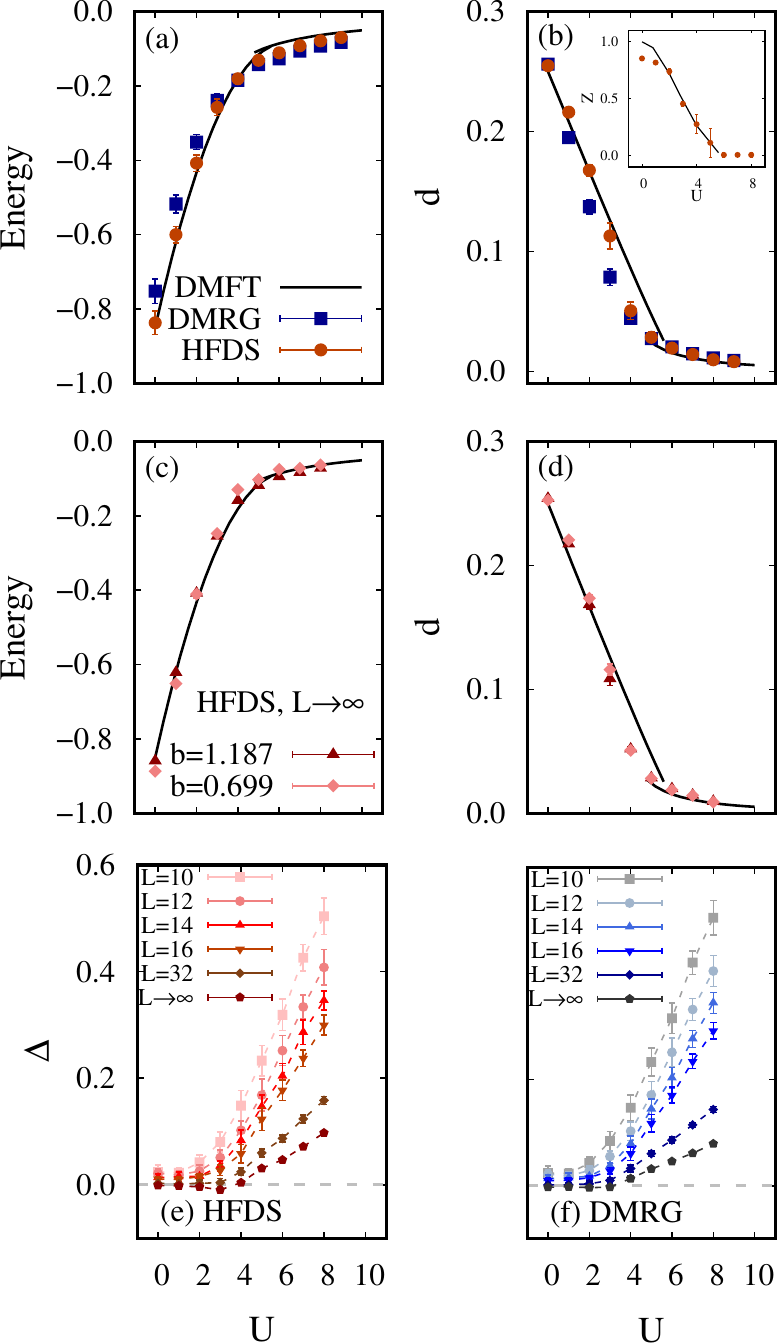}
    \caption{Top panel: Ground state energy per site (a), double occupancy per site (b) and quasi-particle weight (inset b)
    as a function of the Hubbard interaction strength $U$. 
    Results are shown from HFDS (orange circles), DMRG (blue squares) for $L=32$ and DMFT+NRG (black line) 
    calculations for $L=\infty$. Middle panel: Ground state energy (c) and double occupancy (d) obtained from a finite-size extrapolation to $L\rightarrow\infty$ of the HFDS results. The extrapolation is done using the formula $f(L) = A/L^b+f(L\rightarrow\infty)$. Details on the finite-size extrapolations and on the choice of $b$ are available in the SM~\cite{supplemental}. Bottom Panel: Energy gap as a function of $U$ obtained from HFDS (e) and DMRG (f) for various $L$. Details of the \largeL extrapolation are available in \cite{supplemental}.}
    \label{fig:2_Energy_Docc}
\end{figure}

We compare the predictions from the HFDS and DMRG wave function methods to DMFT+NRG data for $L=\infty$ in Fig. \ref{fig:2_Energy_Docc} (a-d). The DMFT+NRG is obtained in a state-of-the-art NRG 
implementation~\cite{Peters_2006, Weichselbaum_2007, Zitko_2009, Lee_2016} based on the QSpace tensor library \cite{Kugler_2022}. In panels (a) and (b) of Figure \ref{fig:2_Energy_Docc}, we plot the ground-state energy per site, the double occupancy, and the quasi-particle weight 
as a function of $U$ for $L=32$.

The results obtained from both methods are overall in good qualitative agreement 
with the DMFT ground-state results, even though the latter apply to the \largeL limit. 
At this system size and values of $U\leq 4$, the accuracy of the HFDS surpasses that of the DMRG, achieving lower variational energies and better agreement with the DMFT double occupancies. 
This noticeable difference persists for increasing system sizes, and in the SM~\cite{supplemental} we compare results for $L = 48$ and show how the DMRG calculation obtains a far higher ground-state energy despite requiring extensive computational resources.
We also note that the DMRG results for the double occupancy are in better agreement with the HFDS and DMFT 
in the insulating phase. This may be expected since this phase has lower entanglement (Fig.~\ref{fig:1_Model_EE}(b)) 
making it more favorable for DMRG in the position basis. Finite-size effects are also expected to be smaller in that 
phase.
In panels (c) and (d) of Figure \ref{fig:2_Energy_Docc}, we display the same observables extrapolated 
to \largeL within the HFDS method (see \cite{supplemental} for details).
The extrapolated values are in good agreement with the DMFT data, 
except for values of $U$ in the range $U\simeq 4$ to $U\simeq 6$, in the vicinity of the Mott transition.
As mentioned above, a metastable insulating solution is known to exist for \largeL in this range of coupling. 
It is unclear whether the observed discrepancy between both the HFDS and DMRG results and the DMFT data in this 
range of $U$ is due to finite-size effects, or if this difference is rather due to the wave function-based 
approaches finding the metastable insulating state.

We also study the ground-state energy gap $\Delta$: 
\begin{equation}
    \Delta = E(N+1) + E(N-1) - 2E(N),
\end{equation}
where $E$ is the ground-state energy averaged over $20$ realizations of the disorder, 
and $N=L$ is the number of particles at half filling.
The results obtained with HFDS and DMRG are reported in panels (e) and (f) of Figure \ref{fig:2_Energy_Docc}, 
respectively, for various sizes $L$. 
An extrapolation to \largeL is also displayed. 

The gap extrapolated to \largeL has a behavior qualitatively similar to the $L=32$ results: 
it is close to $0$ at low values of $U$, and increases above $U=4$, 
the extrapolated value being clearly nonzero for $U=5$ in both DMRG and HFDS. 
Indeed, the gap of the metastable insulator opens at $U_{c1}=4.74\cdots$ in the $L=\infty$ DMFT solution. 
In this limit, however, the ground state remains a metal up to $U_{c2}=5.82\cdots$. 
This again can be interpreted either as the HFDS and DMRG methods picking up a 
metastable insulating solution in the Mott transition region or as much larger system sizes being 
necessary to stabilize the true metallic ground-state for $U\in[U_{c1},U_{c2}]$. 

As a final way of characterizing the Mott transition, 
we show, in the inset of Figure \ref{fig:2_Energy_Docc}(b) the quasiparticle weight $Z$, 
also related to the quasiparticle mass enhancement\cite{Georges_1996},
finding good agreement with the DMFT-computed values of $Z$. 
Details on how we calculate this quantity from the HFDS density matrix are reported in SM~\cite{supplemental}.

\paragraph{Conclusion.} 

Our results demonstrate that wavefunction-based approaches can be used to solve the fully-connected 
random hopping Hubbard model for finite system sizes. 
We showed results for $L=32$ sites, which is far beyond what can be achieved with exact diagonalization. We also 
present, in \cite{supplemental}, computations with $L=48$. 
We find that, for $L=32$ in the metallic regime, the HFDS approach performs better than the DMRG approach. 
The DMRG results we obtain for smaller system sizes ($L\leq 16$) are essentially exact, but the method becomes limited computationally for larger systems and the required resources scale exponentially due to the volume-law entanglement.
The HFDS calculations require comparatively less computational resources than DMRG for the same system size.
In the intermediate coupling regime where metallic and insulating solutions are known from DMFT to coexist in the infinite-volume limit, 
the question of the accuracy of the ground state that is found by our variational calculations is still unclear. 
Calculations for larger system sizes may elucidate this issue. 
We found nonetheless that our calculations are able to clearly identify the Mott transition in this model. 
While DMFT provides an exact, Green's function based, numerical solution of this model in the infinite-volume limit, the nature of its ground-state wave function and how it changes through the Mott transition 
were uncharted territory up to now. 
Our work resolves this by providing a direct determination of this wave function. 

Because the present model captures the essence of the Mott transition, as documented by 
extensions to realistic models of quantum materials in the framework of DMFT~\cite{Georges_1996,Kotliar_2006}, 
our work paves the road for the study of the physics of strong electronic correlations 
with neural quantum states. 
Extensions to closely related models, e.g. of the SYK type~\cite{SYK_review}, 
and to more realistic finite-dimensional models should be within reach in future studies.
Our work furthermore demonstrates that neural quantum states can be used efficiently for the study of 
fermionic problems with volume-law entanglement, in contrast 
to a recent claim~\cite{Passetti_2023} (see also \cite{Zakari_comment_Passetti}). 

\begin{acknowledgments}
\textit{Acknowledgments} - We are most grateful to Fabian Kugler for providing us with the DMFT-NRG data displayed 
in Figure \ref{fig:2_Energy_Docc} and Figure S6 of the SM \cite{supplemental}.
We acknowledge useful discussions with Matthew Fishman, Olivier Parcollet, David Sénéchal, Miles Stoudenmire, Andr\'e-Marie Tremblay and 
Giuseppe Carleo. 
DMRG calculations were done using the ITensor library \cite{ITensor}. Walltimes and RAM usage for the DMRG calculation is based on using a single icelake computing node of the Rusty cluster housed in the Flatiron Institute in New York.
HFDS calculations were done using the Netket library \cite{netket2:2019,netket3:2022}.
The Flatiron Institute is a division of the Simons Foundation. 
\end{acknowledgments}

\bibliography{main.bib}
\clearpage
\section{End Matter}

\subsection{HFDS wave-function amplitudes in the space of electronic configurations}

In this End Matter, we study empirical statistics of the weight of the HFDS wave-function amplitudes for different subsets of electronic configurations and different values of $U$. The aim is to understand the structure of the ground-state wave function and its change as the value of $U$ is increased, changing character from a metal to an insulator. Since the model is disordered, the only distinguishing factor between subsets of electronic configurations is the number of doubly occupied sites $D$ on a given configuration $n$. Consequently, we focus on the analysis of the wave-function amplitudes as a function of $D$ and $U$.

\begin{figure*}
    \centering
    \includegraphics[width=2\columnwidth]{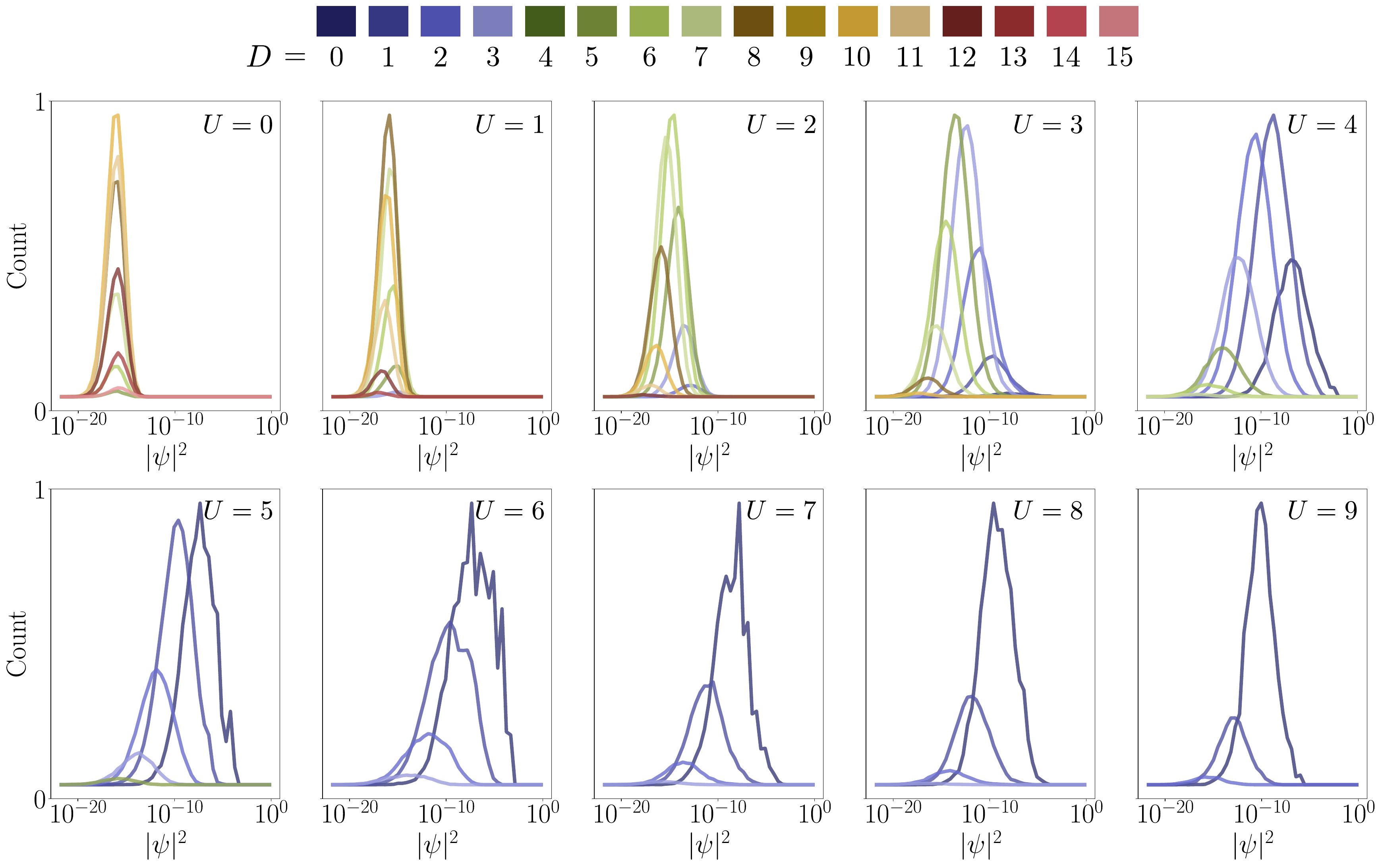}
    \caption{Histograms of the wave-function amplitudes obtained from HFDS for different values of $U$ as indicated in each panel. The wave-function amplitudes are normalized using the norm estimate provided by Eq.~\ref{eq:estimate_norm}. System size is $L=32$. Within each panel, the different colors show the histograms for the amplitudes evaluated at electronic configurations with double occupancy $D$, as indicated by each color.}
    \label{fig:histograms_01}
\end{figure*}

For the largest system size that we consider $(L = 32)$, the direct enumeration of all possible electronic configurations is not feasible, since the dimensionality of the Hilbert space for this system size is $\mathcal{D} = \binom{32}{16}^2 = 3.6(1)\cdot 10 ^{17}$. Therefore, we restrict our statistical study to a subset of $20 \cdot 10^3$ electronic configurations sampled for each $U$ and disorder realization according to the corresponding HFDS ground-state wave-function amplitudes: $n\sim|\psi|^2$. Since the wave-function amplitudes are not required to be explicitly normalized for VMC calculations, the HFDS wave functions are not explicitly normalized. For the study of the wave-function amplitudes as a function of $D$ and $U$, an unbiased estimate of the normalization factor can be approximated from the sampled electronic configurations:
\begin{equation}\label{eq:estimate_norm}
\begin{split}
    \frac{1}{\sum_n |\psi(n)|^2} & = \frac{1}{\mathcal{D}}  \sum_n \frac{1}{\sum_{n'} |\psi(n')|^2} = \frac{1}{\mathcal{D}}  \sum_n \frac{|\bar{\psi}(n)|^2}{|\psi(n)|^2} \\
    & \approx \frac{1}{\mathcal{D}}  \mathbb{E}_{n\sim |\psi|^2}\left[\frac{1}{|\psi(n)|^2} \right],
\end{split}
\end{equation}
where $\bar{\psi}(n)$ labels the explicitly normalized counterpart to $\psi(n)$: $\bar{\psi}(n) = \psi(n) /\sqrt{\sum_{n'} |\psi(n')|^2}$. We utilize this only to plot our results in Fig.~\ref{fig:histograms_01}, although we emphasize that it is an approximation and an accurate, unbiased estimate of the norm is computationally hard \cite{Michal2022}.

Figure~\ref{fig:histograms_01} show the histogram of the normalized wave-function amplitudes for different values of $U$, averaged over 20 disorder realizations. The histograms are grouped by the number of doubly occupied sites $D$ for the sampled electronic configurations. For $U<3$, the weight of the wave function is distributed across electronic configurations with a number of doubly occupied sites around $D \approx 8$. For small $U$, double occupancy does not significantly increase the energy of the system. Since the number of electronic configurations with double occupancy $D$ peaks at $D= 8$, the wave function concentrates around $D = 8$ for small $U$, dominated by a non-interacting character. For $ U = 3 $ and $ U = 4 $, as correlation increases but still in the metallic phase, the weight of the wave function starts shifting from electronic configurations with large numbers of the double occupancy to electronic configurations with $D \in [0, 6]$. In particular, the fraction of sampled configurations with $D = 0$ and $D = 1$ starts to rapidly increase. At $U = 5$, the weight of the wave function is concentrated in electronic configurations with $D \in [0, 3]$. Interestingly, the height of the histogram for $D = 0$ surpasses that of the histogram for $D = 1$. For $U\geq 6$, the weight of the wave function concentrates around electronic configurations with $D = 0$. However, the weight of electronic configurations with $D>0$ is only exactly suppressed in the $U \rightarrow \infty$ limit.

\begin{figure*}
    \centering
    \includegraphics[width=2\columnwidth]{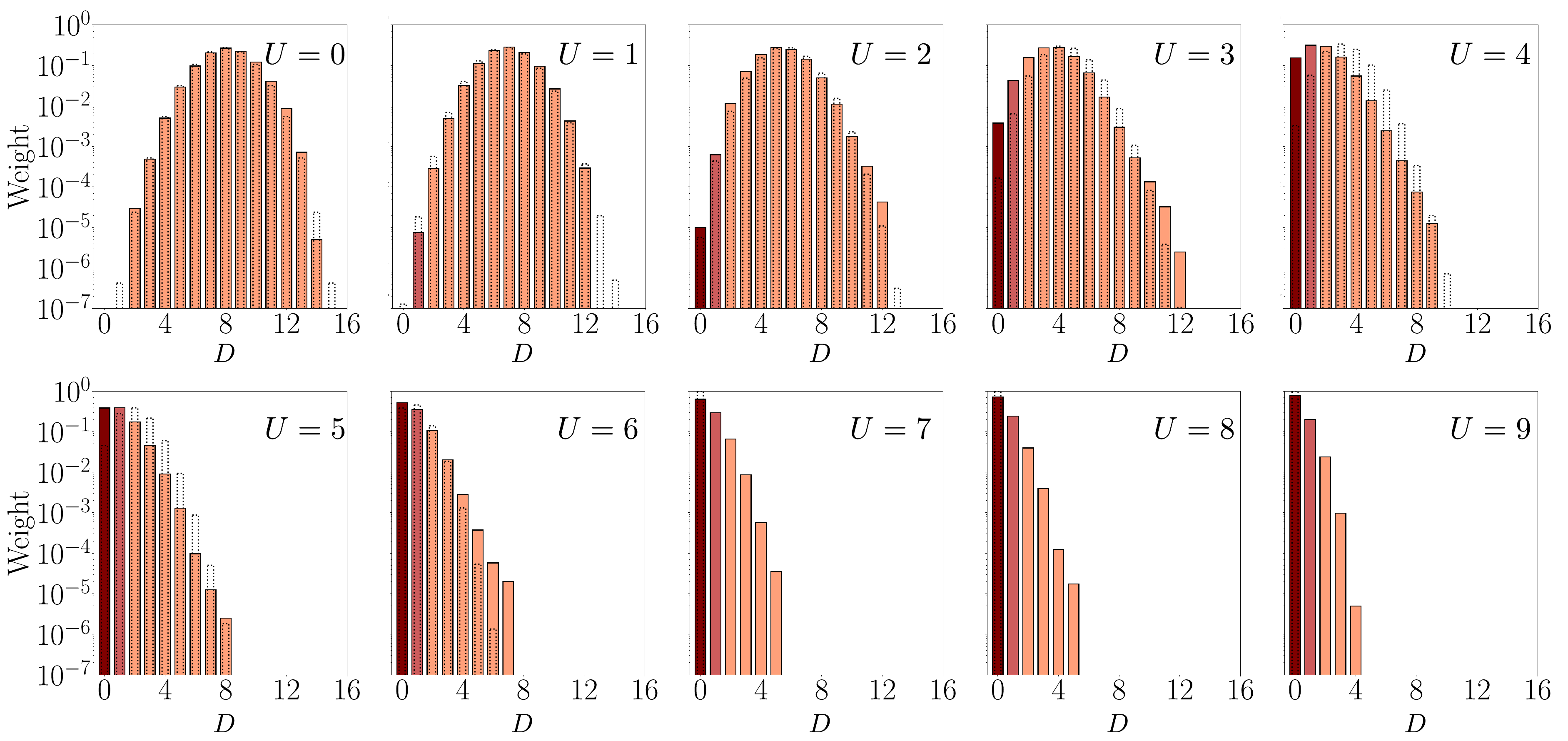}
    \caption{Weights of the HFDS wave function for electronic configurations $n_D$ with double occupancy $D$ as a
    function of $D$. System size is $L = 32$ and different values of $U$ are shown in different panels. Darker to lighter colors indicate $D = 0$, $D = 1$ and $D>1$ respectively. The dotted bars indicate the weights obtained from the Gutzwiller approximation, for reference (see the discussion of Eq.~\ref{Eq:GutzwillerProbabilities}).}
    \label{fig:histograms_02}
\end{figure*}

The information provided by the histograms in Fig.~\ref{fig:histograms_01} can be further summarized by the weight of the wave function for electronic configurations $n_D$ with double occupancy $D$:
\begin{equation}
    \mathcal{W }(D) = \sum_{n_D} \left| \psi(n_D) \right|^2,
\end{equation}
An unbiased estimator for $\mathcal{W}(D)$ can be obtained by the fraction of electronic configurations with double occupancy $D$ sampled according to the probability distribution $P(n_d) \sim |\psi(n_{D})|^2$. Fig.~\ref{fig:histograms_02} shows the weight of the wave function as a function of $D$ for different values of $U$, averaged over twenty disorder realizations. For small values of $U$ in the metallic phase, the wave function spreads across electronic configurations with a wide range of double occupancy. This is expected since there are significantly more electronic configurations with $D>0$, and the energy penalty for having double occupancy is small. Close to the transition point $(U \approx 5-6)$, the weight shifts towards configurations with $D = 0$ and $D = 1$. For large values of $U$ in the Mott insulating phase, the weight of the wave function in each sub-sector of double occupancy decays exponentially with $D$ and is peaked around $D = 0$, while still allowing for charge fluctuations. Additionally, Fig.~\ref{fig:histograms_02} shows the weight of the wave function obtained from the Gutzwiller approximation (see next Section). The predictions form the HFDS ansatz are in good agreement with those from the Gutzwiller approximation deep in the metallic phase. Close to the transition, the Gutzwiller approximation suppresses charge fluctuations more aggressively than the HFDS state. For large values of $U$, the Gutzwiller approximation completely suppresses charge fluctuations, leaving a state with exactly no double occupancy.. 

\clearpage
\setcounter{equation}{0}
\setcounter{figure}{0}   
\renewcommand{\theequation}{S\arabic{equation}}
\renewcommand{\thefigure}{S\arabic{figure}}
\renewcommand{\theHequation}{\theequation}
\renewcommand{\theHfigure}{\thefigure}
\renewcommand{\figurename}{Supplemental Figure}

\section{Supplemental Material to `Mott Transition and Volume Law Entanglement with Neural Quantum States'}

\subsection{Additional insights on the model from DMFT}
Here, we expand on the discussion provided in the main text on the zero-temperature properties of the model known from DMFT 
for $L\rightarrow\infty$. 

For $U=0$, the density of states obeys a Wigner semi-circular law of bandwidth $4t$. 
At finite $U$, the ground-state is a Fermi liquid up to a critical value $U=U_{c2}$. 
The most accurate solutions of the DMFT equations using the numerical renormalization group (NRG) impurity solver report $U_{c2}=5.82...$~\cite{Lee_2017}. 
This model does not host a superconducting ground-state in the repulsive case $U>0$: 
Kohn-Luttinger instabilities to nonlocal order parameters cannot occur, a peculiarity of 
the infinite connectivity. 

For $U>U_{c2}$, the ground-state is a Mott insulator. For any finite $U$, however, charge fluctuations survive 
and the double occupancy remains nonzero. 

The physics of the system in the Mott phase is best understood qualitatively by thinking of the large-$U$ limit. 
There, and as \largeL, the Hamiltonian reduces at half-filling 
to $\sim \frac{4t^{2}}{UL} \left(\sum_i\vec{S}_i \right)^2$ 
which is exactly solvable~\cite{Georges_1993,Rozenberg_1994,Laloux_1994}.
It has an exponentially degenerate manifold of singlet ground-states, 
so that the $T=0$ thermodynamic entropy is 
nonzero and extensive $\sim L\ln 2$. We expect, however, the ground-state to be 
non-degenerate for a finite size $L$ and a given sample $t_{ij}$. Due to the all-to-all hopping, the system is fully frustrated and cannot host a ground-state with N\'eel order.

This model 
provides one of the simplest examples of a system having a $T=0$ phase transition between 
a metallic Fermi liquid and a Mott insulator with no magnetic long-range order. 
The phase transition occurs at $U_{c2}$ and 
is second-order~\cite{Rozenberg_1994,Georges_1996}, 
the energy and double occupancy being continuous through the transition. 
A subtle point is that the insulating solution exists as a metastable, higher energy, solution in a window 
of coupling $U\in [U_{c1},U_{c2}]$, with $U_{c1}=4.74\cdots$~\cite{Lee_2017}.
This holds for the infinite-volume limit, but the consequences for finite-$L$ are not currently known. 

\subsection{Details of DMRG Calculations}

In the DMRG calculations we utilise the ITensor Library to build the long-range Hamiltonian in Eq. \ref{Eq:Hamiltonian} as a MPO and encode our ansatz for the ground state as a MPS. The ordering of the sites is chosen randomly due to the the long-range disordered nature of the Hamiltonian.
\par
\begin{figure}
    \centering
    \includegraphics[width=\columnwidth]{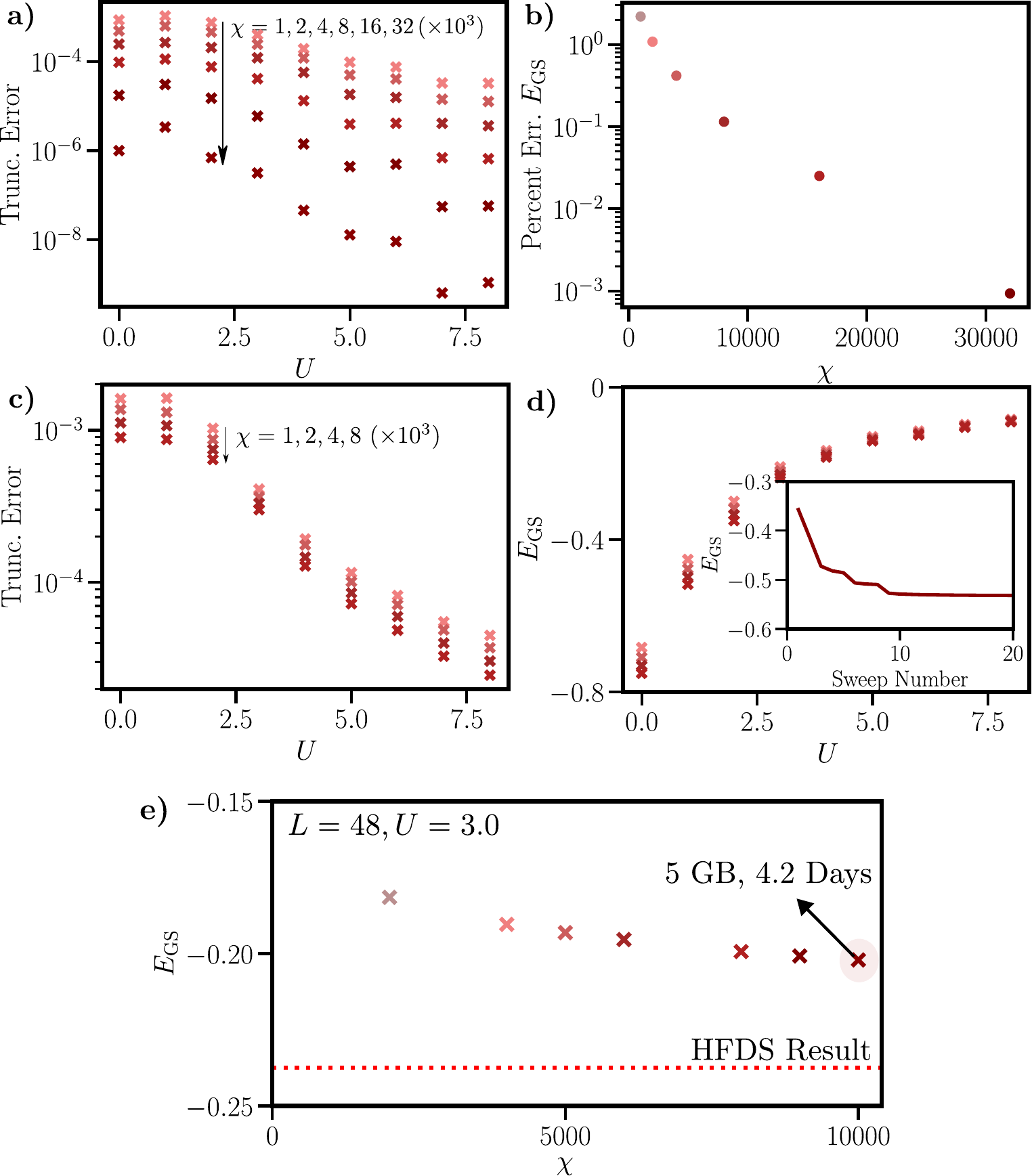}
    \caption{Numerical Details of DMRG calculations for bond-dimensions increasing up to the values used in the main text. a) Average sum of square of singular values thrown away during SVD's in the final sweep of the DMRG calculation. Data is for $L = 16$ averaged over $20$ disorders for a range of $U$ and $\chi$ values. b) Percentage error on the energy found in the DMRG calculation versus $\chi$ and for $L = 16$, $U = 0.0$. c) Average sum of square of singular values thrown away during SVD's in the final sweep of the DMRG calculation. Data is for $L = 32$ averaged over $20$ disorders for a range of $U$ and $\chi$ values. d) Energy of the result of the DMRG calculation for $L = 32$ and a range of $\chi$ and $U$ values. Inset) Convergence of energy with DMRG sweep number for a single disorder instance, $L = 32, U = 1.0$ and $\chi = 8 \times 10^{3}$. e) Energy of the result of the DMRG calculation for $L = 48$, a range of $\chi$ values, $U = 3.0$ and a single disorder instance. The point at $\chi = 10000 $ is annotated with the RAM and walltime required by the simulation. The red dotted line shows the energy obtained from the corresponding HFDS calculation for exactly the same Hamiltonian.}
    \label{fig:S1}
\end{figure}

 In Fig. \ref{fig:S1}a we plot the average truncation error from the SVD's performed in the final (20th) sweep of the calculation for a range of bond-dimensions and for $L= 16$ The results are averaged over the same $20$ instances of disorder as in the main text. A systematic improvement is seen with increasing bond-dimension and for $\chi = 32,000$ the largest error is on the order of $10^{-6}$ for small $U$. We also plot, for $U = 0$, the percentage difference (again averaged over disorders) between the energy density of the state found via DMRG and the exact energy found via diagonalisation of the hopping matrix. For $\chi = 32,000$ the percentage error drops below $0.001 \%$. Given that for $U = 0$ our truncation errors are largest, we are confident this means the energies we have found via DMRG for $L = 16$ throughout the phase diagram are accurate to within $0.001 \%$ or better.
\par We also plot, in Figs. \ref{fig:S1}c and d, the results for $L = 32$. Here, for the largest bond-dimension obtainable ($\chi \sim 8,000$), we reach truncation errors on the order of $10^{-3}$ for small $U$, indicating a significantly more inaccurate calculation than for $L = 16$. This is evidenced by our plots of the convergence of energy versus $U$ and $\chi$, where the ground state energy we find has clearly not converged yet with increasing bond-dimension. We also include an inset for $U = 1.0$, $\chi = 8000$ and a specific instance of disorder. Here we show the change in energy vs DMRG sweep and as we increase the bond-dimension to its maximum value we see this converge with sweep number --- indicating that our DMRG calculations have converged for the given bond dimension.
\par Finally, in Figs. \ref{fig:S1}e, we plot results for $L=48$, $U=3.0$ and a single disorder instance. We show the DMRG found energy (after $20$ sweeps) versus bond dimension and also indicate the result found via HFDS for the same disorder instance, system size and interaction strength. The HFDS result is far below the DMRG result despite the DMRG result requiring over $4$ days of Walltime and $5$GB of RAM on a $64$ core icelake node of our Rusty supercomputing cluster with multithreading enabled.

\subsection{Optimizing the single-particle basis set for DMRG calculations}

\begin{figure}
    \centering
    \includegraphics[width=\columnwidth]{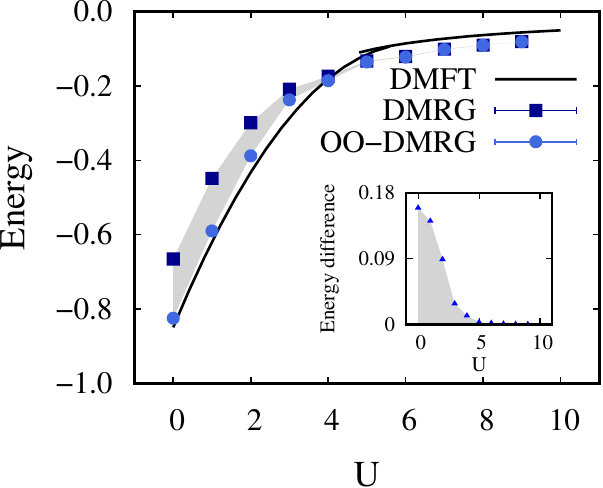}
    \caption{Ground state energy per site as a function of $U$ obtained with DMRG in the fixed localized basis (DMRG) and in an optimized single-particle basis (see text). 
    The system size is $L = 24$ and $\chi_\textrm{max} = 800$ in both cases and the results are averaged over the same $20$ disorder realisations in each case. The shaded region highlights the energy improvement and the inset displays shows the difference in energy between the two approaches as a function of $U$.
    }
    \label{fig:4_OO-DMRG}
\end{figure}

For the sake of completeness, we emphasize that the scaling of the entanglement entropy, 
which is the limiting factor in DMRG, is strongly dependent on the single-particle basis used to represent the Hamiltonian. 
In order to mitigate this effect, we follow the procedure discussed in Ref.~\cite{Robledo2023BasisRotation} to make the DMRG technique invariant under the reparametrization of $H$ corresponding to rotations of the single-particle basis. Every three sweeps, the one- and two-body reduced-density-matrices are measured and gradient descent techniques are used to find the single-particle basis that minimizes the energy. A total of thirty sweeps are performed in total. 
Fig.~\ref{fig:4_OO-DMRG} demonstrates the improvement in the ground state energy resulting from this procedure.
The difference (see inset) between the ground-state energy obtained from DMRG in the fixed localized basis and when including basis optimizations 
is larger in the metallic phase and vanishes in the insulating phase, 
where the ground state is localized and therefore the localized basis is the optimal basis choice for the DMRG. 
This observation can be exploited to obtain additional diagnostics for the Mott transition. 
This will be the focus of future investigations.

\subsection{Details of the VMC calculation with Hidden-Fermion Determinant States}
As described in the main text, VMC is performed within the hidden-fermion formalism, where the basis for the augmented Fock space is spanned by the set of states:

\begin{equation}\label{eq:basis in augmented space}
    |n, \tilde{n} \rangle = \left(\prod_{i, \sigma} \left(\hat{a}^\dagger_{i \sigma} \right)^{n_{i\sigma}}  \right)\left( \prod_{1\leq \mu \leq \tilde{M}} \left(\hat{d}^\dagger_{\mu} \right)^{\tilde{n}_{\mu}} \right)   |0 \rangle,
\end{equation}
where $\hat{a}^\dagger_{i\sigma}$ and $\hat{d}^\dagger_\mu$ are the creation operators for the visible and hidden modes respectively. The previous definition assumes the canonical ordering where the operators closest to the Fock vacuum are those with the largest mode index. VMC in the physical subspace requires the definition of the wave-function amplitudes of the trial state $|\Psi\rangle$ projected into the physical subspace, as described by Eq.~\ref{eq:HFS amplitudes} in the main text.

$|\Psi\rangle$ is chosen as the Slater determinant in the augmented Fock space characterized by a number $N_\textrm{tot} = N + \tilde{N}$ of orbital functions $\phi_n: \{1, \hdots, M_\textrm{tot} \} \mapsto \mathbb{R}$ with $1\leq n \leq N_\textrm{tot}$:

\begin{equation}
    |\Psi\rangle = \hat{\varphi}^\dagger_1 \hdots \hat{\varphi}^\dagger_{N_\textrm{tot}} |0\rangle.
\end{equation}
The new set of creation operators $\hat{\varphi}^\dagger_\alpha$ is a linear combination of the visible and hidden creation operators defined by:

\begin{equation}
    \left(\hat{\varphi}^\dagger_1, \hdots, \hat{\varphi}^\dagger_{N_\textrm{tot}} \right) = \left(\hat{a}^\dagger_1, \hdots, \hat{a}^\dagger_M, \hat{d}^\dagger_1, \hdots, \hat{d}^\dagger_{\tilde{M}} \right) \cdot \Phi.
\end{equation}
$\Phi \in \mathbb{R}^{M_{\textrm{tot}} \times N_{\textrm{tot}}}$ is the matrix of optimizable orbitals, whose columns define the orbital functions. Following Ref.~\cite{Robledo_2022Hidden}, the matrix of orbitals is decomposed into blocks:

\begin{equation}
    \Phi = 
    \begin{bmatrix}
    \phi_{\rm v} & \chi_{\rm v} \\
    \phi_{\rm h} & \chi_{\rm h}
    \end{bmatrix} \enspace .
\end{equation}
Where $\phi_{\rm v} \in \mathbb{R}^{M \times N}$ is the matrix representing the amplitudes of the visible orbitals evaluated in the visible modes, $\chi_{\rm v} \in \mathbb{R}^{M \times \tilde{N}}$ is the matrix representing the amplitude of the hidden orbitals evaluated in the visible modes, $\phi_{\rm h} \in \mathbb{R}^{\tilde{M} \times N}$ is the matrix representing the amplitude of the visible orbitals evaluated in the hidden modes and $\chi_{\rm h} \in \mathbb{R}^{\tilde{M} \times \tilde{N}}$ is the matrix representing the amplitude of the hidden orbitals evaluated in the hidden modes.

In the VMC framework particle configurations ($x = (x_1, \hdots, x_N)$ and $\tilde{x} = (\tilde{x}_1, \hdots, \tilde{x}_{\tilde{N}})$ for visible and hidden fermions respectively) are sampled instead of occupations. $|x, \tilde{x}\rangle$ states differ from $|n, \tilde{n} \rangle$ states in that the application of the creation operators to the Fock vacuum does not follow the canonical order of Eq.~\ref{eq:basis in augmented space} and instead follows the order specified by the particle index labels in $x$ and $\tilde{x}$. The constraint used to project the trial state to the physical subspace $F(n)$ can be translated into a constraint function in the space of particle configurations:

\begin{equation}
    f:x\mapsto \tilde{x},
\end{equation}
where $f$ is chosen to be symmetric with respect to the particle indices of $x$. This constraint ensures that the Fermi statistics is preserved. The hidden-fermion determinant state wave-function amplitudes in the space of particle configurations are then given by:

\begin{equation}
     \psi(x) = \langle x, f(x) |\Psi \rangle =
    \det
    \begin{bmatrix}
    \phi_{\rm v}(x)    & \chi_{\rm v}(x) \\
    \phi_{\rm h}(f(x)) & \chi_{\rm h}(f(x))
    \end{bmatrix}.
\end{equation}
$\big[ \phi_{\rm v}(x)$, $\chi_{\rm v}(x)\big ]$ and $\big[ \phi_{\rm h}(f(x)), \chi_{\rm h}(f(x))\big]$ denote the $N\times(N+\tilde{N})$ and $\tilde{N}\times(N+\tilde{N})$ sub-matrices obtained from $\big[\phi_{\rm v}, \chi_{\rm v}\big]$ and $\big[ \phi_{\rm h}, \chi_{\rm h}\big]$ respectively by slicing the row entries corresponding to $x$ and $f(x)$. The $\big[ \phi_{\rm h}(f(x)), \chi_{\rm h}(f(x))\big]$ matrix is denoted as the \textit{hidden sub-matrix}.

Following the prescription of Ref.~\cite{Robledo_2022Hidden}, the hidden sub-matrix can be materialized by the direct parametrization of the composition of the functions $\phi_\textrm{h}(f(x))$ and $\chi_\textrm{h}(f(x))$ by neural networks, whose input is the occupancy $n$ associated to the particle configuration $x$. This choice of parametrization automatically satisfies the permutation-symmetry requirements of $f(x)$. In practice, each of the $\tilde{N}$ rows of the hidden sub-matrix is parametrized by a two-layer perceptron with hyperbolic tangent activation functions. The number of hidden units in the hidden layer is characterized by the ratio $\alpha$ between the number of hidden units and the dimensionality of the input space ($2L$ in this case).

The complete set of variational parameters is given by the matrices $\phi_\textrm{v}$ and $\chi_\textrm{v}$ together with the weights and biases of the neural networks that parametrize the hidden sub-matrix. The hyper-parameters that control the expressive power of the ansatz are the number of hidden fermions $\tilde{N}$ and the width of the neural networks $\alpha$.

The wave function is optimized using the stochastic reconfiguration parameter update rule~\cite{Sorella1998StochasticReconfiguration, sorella2007StochasticReconfiguration}. The stochastic reconfiguration is the application of the natural gradient descent technique~\cite{amari1998natural} to the optimization of quantum states~\cite{Stokes2020quantumnatural}.

We observe that convergence to states with lower energies can be achieved if the optimization of the variational parameters is performed in two steps. First, the rows of the $\big[\phi_\textrm{v}, \chi_\textrm{v}\big]$ matrix are fixed to be the $N+ \tilde{N}$ Hartree-Fock orbitals of lowest energy (the Hartree-Fock orbitals are obtained using the PySCF~\cite{PySCF_1, PySCF_2, PySCF_3} library), while the weights and biases of the hidden sub-matrix are optimized for a fixed number of iterations. This produces a compact representation of a wave function that contains the linear combination of all possible single, double, ..., $\tilde{N}$-tuple excitations to the lowest $\tilde{N}$ virtual Hartree-Fock orbitals. Each factor of the linear combination carries a Jastrow factor determined by the hidden sub-matrix coefficients (see Ref.~\cite{Robledo_2022Hidden} for details). In the second step, both $\big[\phi_\textrm{v}, \chi_\textrm{v}\big]$ are optimized with the weights and biases of the neural networks in the hidden sub-matrix.

\subsection{Gutzwiller Results}
\label{app:Gutzwiller}
We also computed the results for the wave function amplitudes predicted by the Gutzwiller ansatz \cite{Vollhardt1984}. The Gutzwiller ansatz is an unnormalized ansatz for the ground state of the Hubbard model which reads
\begin{equation}
    \ket{\psi} = \prod_{i}\left (1 - (1-g)\hat{n}_{i, \uparrow}\hat{n}_{i, \downarrow} \right)\ket{\psi_{0}},
\end{equation}
where $g$ is a variational parameter and $\ket{\psi_{0}}$ is a reference state which is typically taken to be the ground state of the non-interacting ($U = 0$) model. We choose to use such a reference state here. In the half-filled case we focus on here the ansatz predicts the double occupancy
\begin{equation}
    d_{i} = \langle \hat{n}_{i, \uparrow}\hat{n}_{i, \downarrow} \rangle = \begin{cases}
    \frac{1}{4}\left(1 - \frac{U}{8E_{0}}\right) & U < 8E_{0} \\
    0 & {\rm otherwise}
    \end{cases}
    \label{Eq:GutzwillerDoubleOccupancy}
\end{equation}
where $E_{0}$ is the non-interacting ground state energy density. For any finite-size system the ansatz thus predicts a simple linear dependence of the double occupancy vs $U$ in the metallic regime and a vanishing double occupancy throughout the metallic regime. This is not the same as our HFDS and DMRG results which go beyond the Gutzwiller ansatz and correctly show a breakdown of this linear dependence in the crossover regime and an asymptotic approach to $D = 0$ in the metallic regime. 
\par With the Gutzwiller ansatz we can also predict the weight of the wave function in each of the double occupancy sectors. Specifically, given $g$, we have
\begin{equation}
    P(D) = \sum_{n \in D}\vert \psi(n) \vert^{2} = \frac{g^{2D} L! }{4^{L}(\frac{L}{2} - D!)^{2}(D!)^{2}},
    \label{Eq:GutzwillerProbabilities}
\end{equation}
for the probability of finding the wave function in a configuration with double occupancy $D$ -- with the sum taken over all configurations with $D$ doublons. We can ascertain the value for $g$ for a given $U$ and thus compute $P(D)$ by combining Eqs (\ref{Eq:GutzwillerDoubleOccupancy}) and (\ref{Eq:GutzwillerProbabilities}) to get
\begin{equation}
    \frac{\sum_{D = 0}^{L/2}\frac{Dg^{2D} }{(\frac{L}{2} - D!)^{2}(D!)^{2}}}{\sum_{D = 0}^{L/2}\frac{g^{2D}  }{(\frac{L}{2} - D!)^{2}(D!)^{2}}} = \begin{cases}
    \frac{1}{4}\left(1 - \frac{U}{8E_{0}}\right) & U < 8E_{0} \\
    0 & {\rm otherwise}
    \end{cases}
\end{equation}
which is a polynomial in $g$ and thus $g$, and consequently $P(D)$, can be found by finding the real root of the polynomial with $0 \leq g \leq 1$. In Fig. \ref{fig:histograms_02} of the End Matter, we compare the wave-function amplitudes of the Gutzwiller ansatz to those predicted by HFDS. We see a strong similarity between the two deep in the metallic phase where Gutzwiller is most accurate. As the critical region is approached however, the Gutzwiller ansatz gives a much larger weight to the $D = 0$ sector than HFDS --- eventually predicting that the entirety of the wave function lies in the $D = 0$ sector beyond the critical point. HFDS thus goes beyond the Gutzwiller ansatz and provides more accurate insights into the nature of the metal-insulator crossover.

\subsection{Quasi-particle weight}
\label{app:quasiPartWeight}
\begin{figure*}
    \centering
    \includegraphics[width=2\columnwidth]{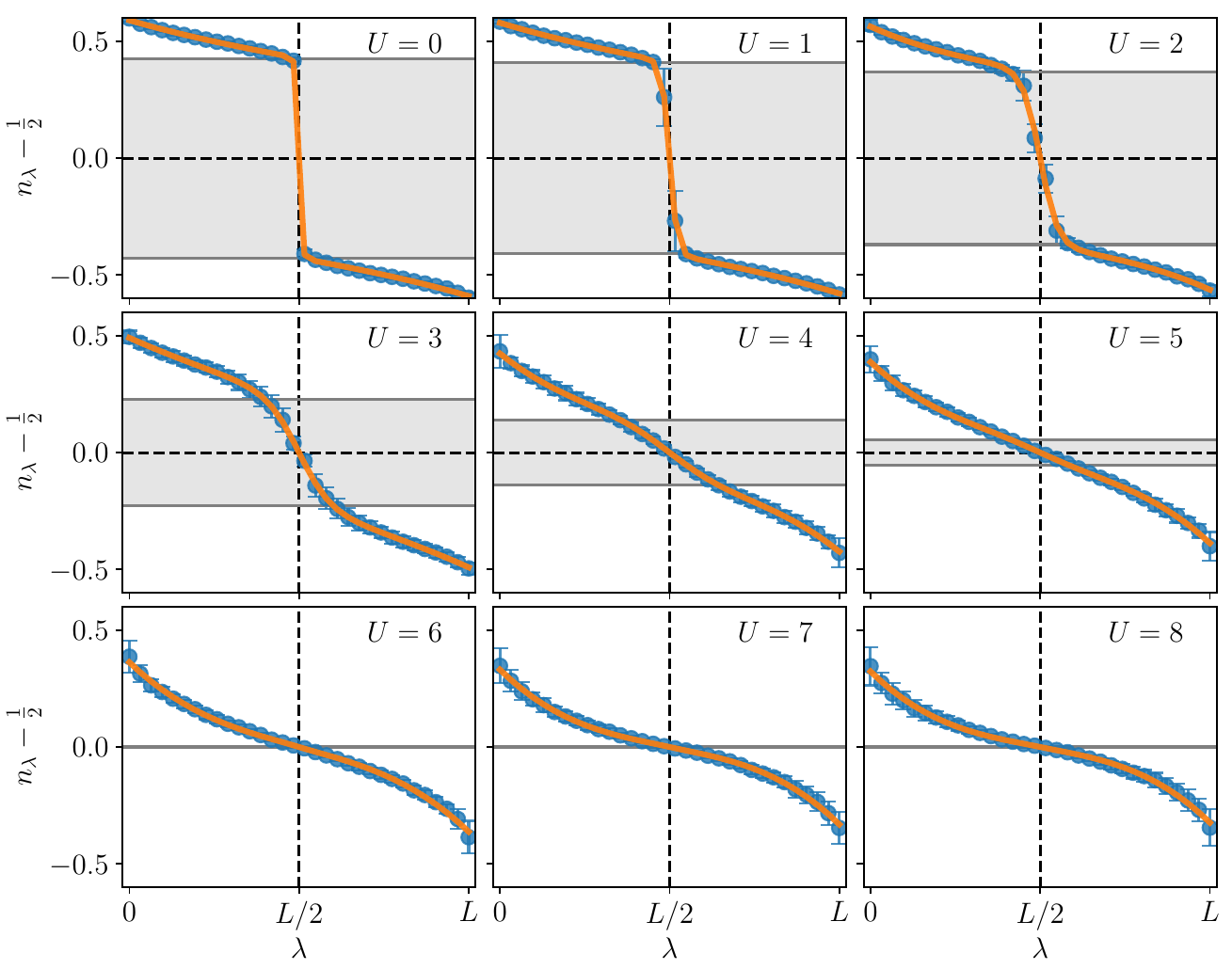}
    \caption{Eigenvalues of the density matrix $n_\lambda$ as a function of $\lambda$. Different panels corresponds to different values of $U$ as indicated. Blue dots are the values obtained from HFDS for $L = 32$ sites, averaged over twenty disorder realizations. The orange curves are the fit to the functional form in Eq.~\ref{eq:fitting_n_lambda}. The dashed vertical line is centered at the Fermi level. The dashed horizontal line is for visual reference at $n_\lambda-1/2 = 0$. The shaded grey area represents the amplitude of the sigmoid in the fitting function (fitting parameter $a$), which corresponds to the quasi-particle weight $Z$.}
    \label{fig:n_lambda}
\end{figure*}

The Mott transition may also be characterized by the quasi-particle weight $Z$. It can be obtained from the density 
matrix:
\begin{equation}
    \rho_{ij} = \frac{1}{2}\sum_{\sigma} \langle c^{\dagger}_{i\sigma}c_{j\sigma} \rangle.
    \label{eq:rho_ij}
\end{equation}
We compute $\rho_{ij}$ for a given disorder realization and obtain its eigenvalues $n_\lambda$. The eigenvalues are averaged over disorder realizations. Fig.~\ref{fig:n_lambda} shows $\overline{n_\lambda}$ for a system with $L = 32$ sites obtained from HFDS. At small $U$ we observe that some of the values of $\overline{n_\lambda}$ are outside the allowed values of $\overline{n_\lambda} \in [0, 1]$. We attribute this discrepancy to the statistical error coming from the Monte Carlo estimation of $\rho_{ij}$, which is then propagated in the diagonalization procedure of the density matrix. Likewise, we also observe that at $U = 0$, $\overline{n_\lambda}$ is not a perfect step function. We again attribute this discrepancy to the statistical error in the calculation of $\rho_{ij}$ propagating in the diagonalization procedure.

The quasi-particle weight $Z$ is defined as the discontinuity of $n_\lambda$ at the Fermi level, which in this case corresponds to $\lambda = L/2$. Finite-size effects ``soften'' the discontinuity in $n_\lambda$ (see Fig.~\ref{fig:n_lambda}) and therefore, the difference $n_{L/2-1}-n_{L/2+1}$ does not provide an accurate estimate for $Z$.

Alternative, the amplitude of smeared out jump in $\overline{n_\lambda}$ can be captured by fitting the spectra of the density matrices in Fig.~\ref{fig:n_lambda} to the functional form:
\begin{equation}\label{eq:fitting_n_lambda}
    f(\lambda) = a \cdot \left(\frac{1}{1+e^{b \cdot (\lambda-L/2)}}- \frac{1}{2} \right) - c \cdot \lambda -d \cdot \lambda^3,
\end{equation}
where $a$, $b$, $c$ and $d$ are fitting parameters. The term proportional to $a$ is a sigmoid function representing the smeared out jump, of amplitude $a$ and width inversely proportional to $b$. Thus, the value of $Z$ can be obtained directly from the value of the fitting parameter $a$. The linear and cubic terms in $f(\lambda)$ are included to fit the other features in the density-matrix spectrum observed in Fig.~\ref{fig:n_lambda}. Fig.~\ref{fig:n_lambda} shows that the proposed functional form in Eq.~\ref{eq:fitting_n_lambda} provides a very accurate fit to the computed values of $\overline{n_\lambda}$ from HFDS.

\begin{figure}
    \centering
    \includegraphics[width=1\columnwidth]{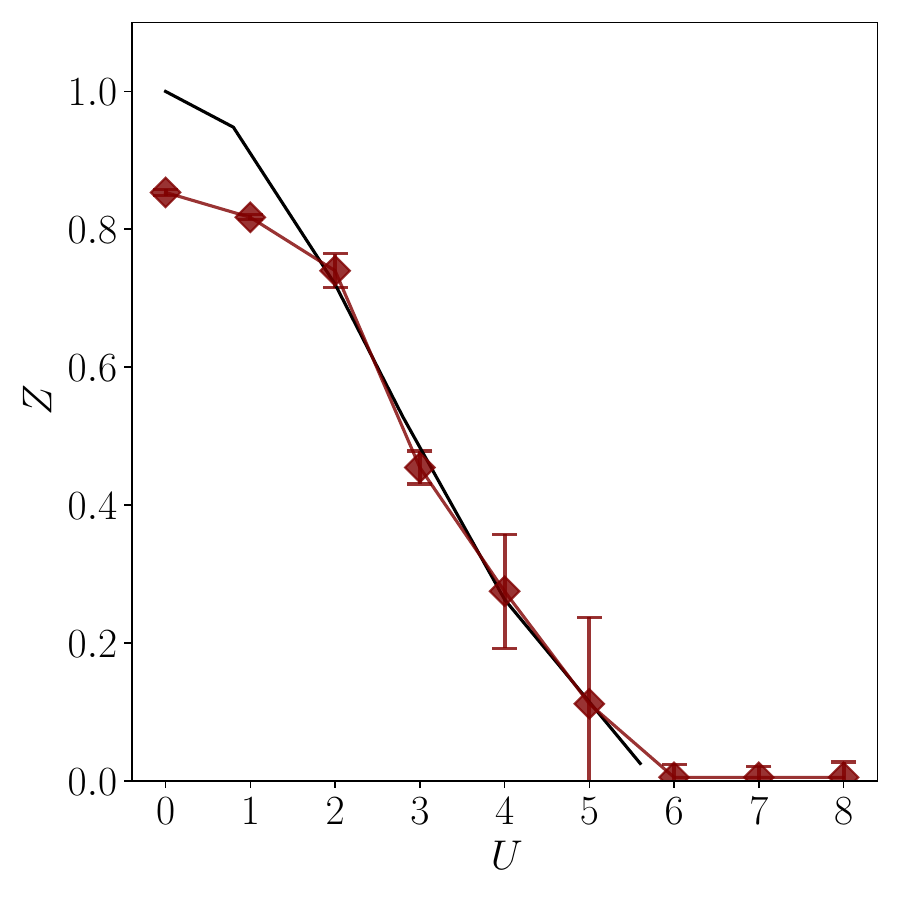}
    \caption{Expanded from the inset Fig. \ref{fig:2_Energy_Docc} (b) of the main text. Weight $Z$ as a function of $U$. Maroon diamonds are the values obtained from HFDS for $L = 32$, obtained from the fittings shown in Fig.~\ref{fig:n_lambda} (fitting parameter $a$ in Eq.~\ref{eq:fitting_n_lambda}). The error bars correspond to the fitting error for the parameter $a$. The black solid line is the DMFT result in the $L \rightarrow \infty$ limit. }
    \label{fig:Z}
\end{figure}

The quasi-particle weight $Z$ extracted from the fits in Fig.~\ref{fig:n_lambda} is shown in Fig.~\ref{fig:Z} and included in Fig.~\ref{fig:2_Energy_Docc} of the main text. The DMFT value for $Z$ in the $L \rightarrow \infty$ limit is also included for reference. For $U>1$ the agreement between the HFDS and DMFT value for $Z$ is remarkable. The vanishing of $Z$ for $U \approx 5-6$ clearly marks the metal to Mott insulator transition. This observation clearly shows that the HFDS are capable of accurately describing many-body phenomena to a large degree of accuracy. The discrepancy in the value of $Z$ between DMFT and HFDS at $U = 0$ and $U = 1$ are a direct consequence of the statistical errors in the computation of the matrix elements of the density matrix, that propagate though the diagonalization procedure, as mentioned above.

\subsection{Finite-size extrapolation to infinite system size}
\label{app:finite_size}

Due to computational limitations, we perform the DMRG and HFDS calculations with systems of finite size $L$. The collected results for various values of $L$ can then be used to obtained extrapolated results at $L\rightarrow\infty$, as shown in Figure~\ref{fig:2_Energy_Docc}. Here, we detail how such extrapolations are performed.

\begin{figure}
    \centering
    \includegraphics[width=0.9\columnwidth]{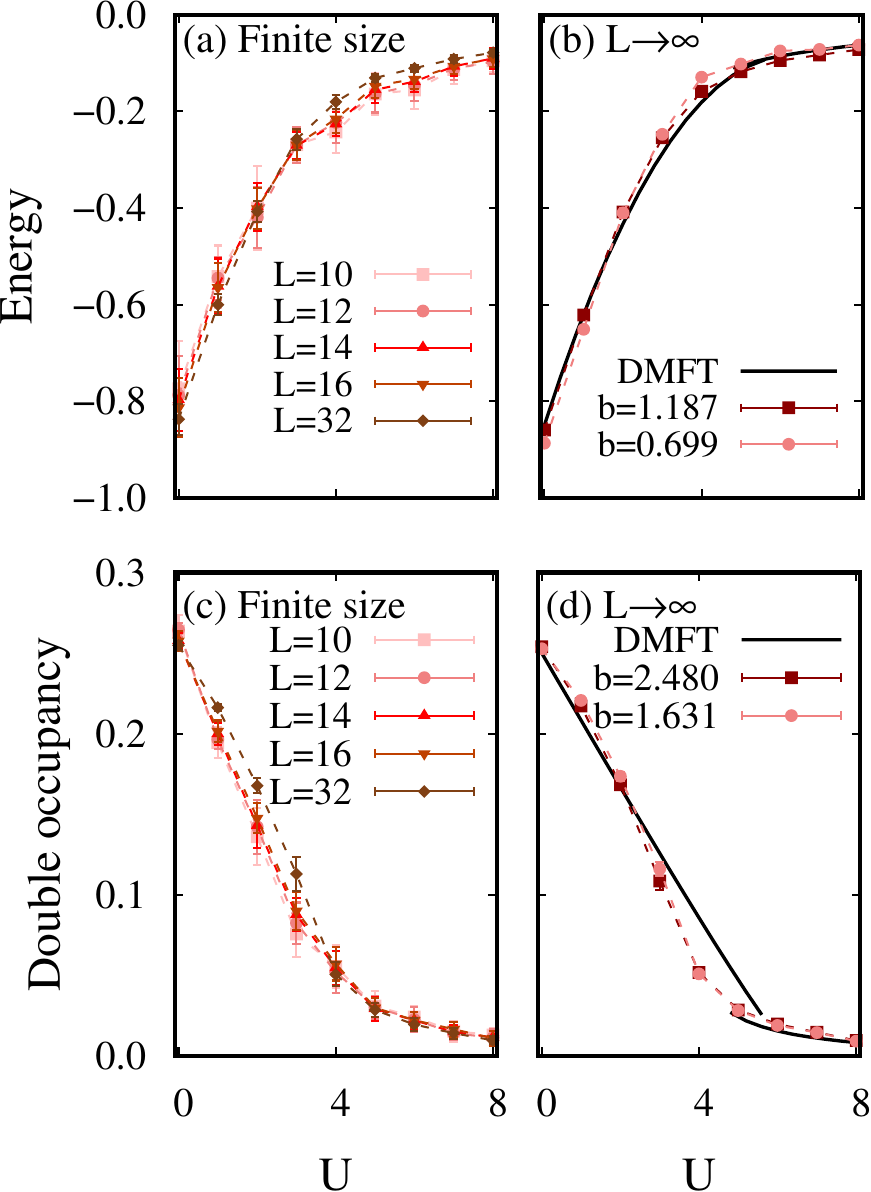}
    \caption{Finite-size scaling of the ground state energy and double occupancy as a function of the Hubbard interaction strength $U$ from HFDS calculations. Panels (a) and (c) show the finite-size results of the energy and double occupancy respectively, for system sizes $L=10,~12,~14,~16$ and $32$. Panels (b) and (d) show the extrapolations to infinite dimensions $L\rightarrow\infty$ using the formula $f(L) = A/L^b+f(L\rightarrow\infty)$, where the exponent $b$ takes the values specified in the figures.}
    \label{fig:finite_size_extrap_Energy_Docc_HFDS}
\end{figure}

We assume that the dependence on the system size of the quantities of interest follows the equation
\begin{equation}
    f(L) = A/L^b+f(L\rightarrow\infty),
    \label{eq:fit_scaling}
\end{equation}
where the extrapolated quantity of interest is $f(L\rightarrow\infty)$. In the case of the ground state energy and double occupancy, we first perform a fit of the HFDS data obtained in the metallic regime at $U=0$ with equation \ref{eq:fit_scaling} and leave all the fit parameters free. This yielded the exponent $b=1.187$ for the energy and $b=2.480$ for the double occupancy. We then perform a second fit in the insulating regime ($U=6$ for the energy, and $U=8$ for the double occupancy), from which we obtain the exponents $b=0.699$ for the energy and $b=1.631$ for the double occupancy. Next, we fix the value of $b$ to the values detailed above in the fit function \ref{eq:fit_scaling} and perform the fits for the energy and the double occupancy at all values of $U$. The results for the extrapolated quantities at $L\rightarrow\infty$ are shown in panels (b) and (d) of Figure \ref{fig:finite_size_extrap_Energy_Docc_HFDS}. Ultimately, the choice of the exponent does not change significantly the qualitative and quantitative behavior of the extrapolated quantities.

In the case of the energy gap, we manually choose the value of the exponent $b=1.65$ as it yields relatively smooth results at all values of $U$ for both the HFDS and DMRG calculations.

\end{document}